\documentclass{article}
\usepackage{graphicx}
\graphicspath{%
    {converted_graphics/}
    {C:/1KGLaaa/aatempfold/Papers/SDAK/}
}
\begin{document}

\begin{center}
{\LARGE An Edge-Enhancing Crystal Growth Instability}\vskip6pt

{\LARGE Caused by Structure-Dependent Attachment Kinetics}\vskip6pt

{\Large Kenneth G. Libbrecht}\vskip4pt

{\large Department of Physics, California Institute of Technology}\vskip-1pt

{\large Pasadena, California 91125}\vskip-1pt

\vskip18pt

\hrule\vskip1pt \hrule\vskip14pt
\end{center}

\textbf{Abstract.} We describe a novel crystal growth instability that
enhances the development of thin edges, promoting the formation of
plate-like or hollow columnar morphologies. This instability arises when
diffusion-limited growth is coupled with structure-depdendent attachment
kinetics, specifically when the nucleation barrier on a facet surface
decreases substantially as the facet width approaches atomic dimensions.
Experimental data are presented confirming the presence of this instability
in the growth of ice from water vapor at -15 C. We believe this
edge-enhancing effect plays an important role in determining the growth
morphologies of atmospheric ice crystals as a function of temperature, a
phenomenon that has been essentially unexplained for over 75 years. Our
model of structure-dependent attachment kinetics appears to be related to
surface melting, and thus may be present in other material systems, whenever
crystal growth from the vapor phase occurs near the material melting point.

\section{Introduction}

The formation of crystalline structures during solidification yields a
fascinating variety of morphological behaviors, resulting from the sometimes
subtle interplay of non-equilibrium surface processes at the molecular
scale. In many cases, seemingly small changes in molecular dynamics at the
nanoscale can produce large morphological changes at all scales. Some
examples include free dendritic growth from the solidification of simple
liquids, where slight anisotropies in the interfacial surface energy
determine the overall characteristics of the growth morphologies \cite%
{dendrites, brener}, whisker growth from the vapor phase initiated by single
screw dislocations and other effects \cite{whiskers}, the formation of
porous aligned structures from directional freezing of composite materials 
\cite{directional}, and a range of other pattern formation processes \cite%
{cross, book}. Since controlling structure formation during solidification
has application in many areas of materials science, much effort has been
directed toward better understanding the underlying physical effects and
their interactions.

One oft-studied example is the formation of ice crystals from water vapor in
an inert background gas. Although this is a relatively simple physical
system, ice crystals exhibit a remarkable variety of columnar and plate-like
forms, and much of the phenomenology of their growth remains poorly
understood \cite{libbrechtreview}. Because ice plays important roles in many
environmental and biological processes, understanding the detailed molecular
structure and dynamics of the ice surface has received significant attention 
\cite{clouds, moldym1, dash}. The present study was undertaken to use ice
crystal growth from water vapor as essentially an experimental probe of the
ice surface dynamics under changing conditions. In this regard we are using
ice as a convenient test crystal, with our overarching goal being to better
understand the molecular dynamics that govern crystal growth behaviors more
generally.

Observations of ice crystal growth from water vapor dating back to the 1930s 
\cite{libbrechtreview, nakaya} have revealed a complex and puzzling
morphological dependence on temperature. Under common atmospheric
conditions, for example, ice crystals typically grow into thin plate-like
forms near -2 C, slender columns and needles near -5 C, thin-walled hollow
columns near -7 C, very thin plates again near -15 C, and columns again
below -30 C. These observations are often summarized in the well-known
Nakaya morphology diagram \cite{libbrechtreview}, and for over 75 years this
remarkable behavior as a function of temperature has remained essentially
unexplained. After considerable research effort, the underlying physical
mechanisms that govern the morphologies of growing ice crystals remain quite
poorly understood. \cite{libbrechtreview, nelson, pruppacher}.

The temperature dependence seen in ice growth is likely related to
temperature-dependent surface melting, but even this is not known with
certainty. Surface melting describes an equilibrium structural organization
of the crystal surface, and its temperature dependence is itself only poorly
understood \cite{dash}. How surface melting in turn affects a highly
dynamical process like crystal growth is clearly a very difficult problem.

While investigating this phenomenon, we have come to realize that
diffusion-limited growth coupled with the usual assumptions regarding the
water vapor surface attachment kinetics cannot explain the growth of thin
plate-like ice crystals. To solve this dilemma and explain the disparate
growth measurements, we recently proposed that the attachment kinetics does
not depend solely on temperature and supersaturation, as is typically
assumed, but depends also on the morphological structure of the crystal. We
refer to this phenomenon as \textit{structure-dependent attachment kinetics}
(SDAK) \cite{sdak}.

In this paper we better define the SDAK model, show that it leads to an
edge-enhancing growth instability, and we describe experimental data and
crystal growth modeling that strongly support this hypothesis. For the ice
case, the evidence suggests that SDAK effects play a substantial role in
determining growth morphologies, especially the formation of thin plate-like
and hollow columnar crystals. For this reason, we believe the SDAK model
represents a significant step toward understanding the physical basis of the
Nakaya morphology diagram. Moreover, we suggest that the SDAK effect and its
resulting growth instability may have more general application in crystal
growth dynamics, especially when crystal growth occurs in the presence of
significant surface melting.

\section{The SDAK Model}

\subsection{Intrinsic Growth of the Principal Facets}

\begin{figure}[tb] 
  \centering
  \includegraphics[width=3.2in,keepaspectratio]{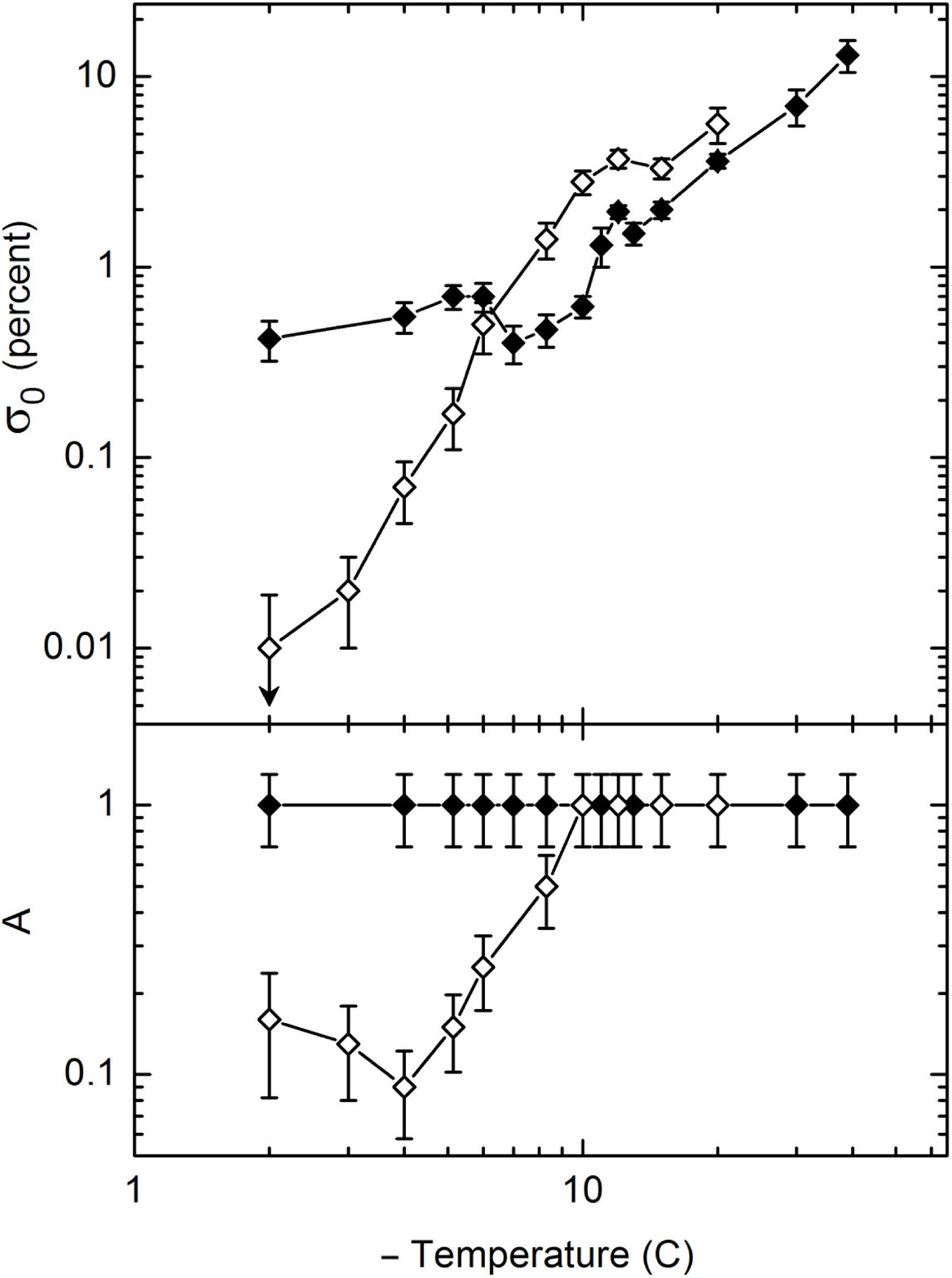}
  \caption{Measurements of the intrinsic
growth rates of the principal ice crystal facets. The growth velocity normal
to the surface is described by $v=\protect\alpha v_{kin}\protect\sigma %
_{surf}$, where $\protect\sigma _{surf}$ is the supersaturation at the
surface and the attachment coefficient is parameterized with $\protect\alpha %
(T,\protect\sigma _{surf})=A\exp (-\protect\sigma _{0}/\protect\sigma %
_{surf}).$ The solid points show the measured $A(T)$ and $\protect\sigma %
_{0}(T)$ for the basal facets, while the open points show measurements of
the prism facets, from \protect\cite{intrinsic}.}
  \label{intrinsic}
\end{figure}

A first step toward understanding
ice crystal growth dynamics from water vapor is to quantify the growth rates
of the principal facet surfaces. Following the notation in \cite%
{libbrechtreview}, we parameterize the surface growth velocities using $%
v=\alpha v_{kin}\sigma _{surf},$where $v$ is the perpendicular growth
velocity, $v_{kin}(T)$ is a temperature-dependent \textquotedblleft
kinetic\textquotedblright\ velocity derived from statistical mechanics, and $%
\sigma _{surf}$ is the water vapor supersaturation relative to ice at the
growing surface. The attachment coefficient $\alpha ,$ which depends on $T$, 
$\sigma _{surf},$ and other factors, encapsulates the molecular dynamics
governing crystal growth at the crystal/vapor interface. From the definition
of $v_{kin}$ we must have $\alpha \leq 1.$

For the simplest case -- the growth of an infinite, clean, dislocation-free
faceted ice surface in near equilibrium with pure water vapor at a fixed
temperature -- the attachment coefficient is well defined and we must have a
unique $\alpha (\sigma _{surf},T)$ for each facet surface. We refer to the $%
\alpha (\sigma _{surf},T)$ in this ideal case as the \textquotedblleft
intrinsic\textquotedblright\ attachment coefficients for a given surface.

We determined $\alpha (\sigma _{surf},T)$ for the prism and basal facets
through a lengthy series of measurements of crystals growing on a substrate
at low background pressure. Experimental details, the resulting data and
analysis, and references to prior work measuring $\alpha (\sigma _{surf},T),$
can be found in \cite{intrinsic}. Over the temperature range $-2$ C $>T>-40$
C, and for both facets, our growth data are well described by a
dislocation-free layer-nucleation crystal growth model, which we
parameterize as $\alpha (\sigma _{surf})=A\exp (-\sigma _{0}/\sigma _{surf})$%
. The measured parameters $A(T)$ and $\sigma _{0}(T)$ for the basal and
prism facets are shown in Figure \ref{intrinsic}.

Note that our understanding of the detailed molecular structure and dynamics
of the ice surface is not sufficient to provide an explanation for these
data. The functional form for $\alpha (\sigma _{surf})$ comes from classical
nucleation theory, and this theory dictates that the parameter $\sigma _{0}$
is related to the step energy $\beta $ associated with the edge of a
molecular terrace on the facet surface \cite{intrinsic}. But the temperature
dependence of $\alpha (\sigma _{surf})$ is likely related to the details of
surface melting on the two facets, for which there is only the most
rudimentary theoretical description \cite{dash}. Thus for the present
discussion we simply accept the measured $A(T)$ and $\sigma _{0}(T)$ in
Figure \ref{intrinsic} as empirical fact.

\subsection{Structure Dependent Attachment Kinetics}

The measured intrinsic growth rates, as parameterized above, immediately
appears to contradict many details in the Nakaya morphology diagram. For
example, from Figure \ref{intrinsic} we see that $\sigma _{0,basal}<\sigma
_{0,prism}$ at -15 C, implying that $\alpha _{prism}<\alpha _{basal}$ for
all supersaturations at this temperature. This inequality suggests that
columnar prisms would be the preferred growth morphology, while it is well
established that thin plates form at this temperature.

The SDAK model was created to explain this discrepancy, suggesting that the
attachment coefficient $\alpha (\sigma _{surf})$ can depend on the mesoscale
morphological structure of the ice surface itself. In particular, our SDAK
model at -15 C assumes that $\alpha _{prism}$ on a thin plate edge is higher
than the corresponding \textquotedblleft intrinsic\textquotedblright\ $%
\alpha _{prism}$ for a large faceted surface. The increased $\alpha _{prism}$
then reverses the above inequality, yielding $\alpha _{prism}>\alpha
_{basal} $ and resulting in the growth of thin plates.

We now carry this model one step further and propose that the increase in $%
\alpha _{prism}$ on the edge of a thin plate is caused by a reduction of the
nucleation barrier on that surface, specifically a reduction of $\sigma
_{0,prism}$. More generally, we put forth the hypothesis that $\sigma
_{0,prism}$ decreases as the width of the final molecular terrace on the
prism surface decreases. The same SDAK effect may be present on the basal
facet as well, and its relative importance on both facets will be determined
by detailed surface dynamics. After examining the physical motivation for
the SDAK model in more detail, along with its subsequent ramifications, we
then treat it as a testable hypothesis for comparison with experimental data.

\subsection{SDAK Microscopic Model}

As a qualitative justification for the SDAK hypothesis, consider the
molecular structure and dynamics on the edge of a thin plate crystal. The
radius of curvature of the edge is perhaps $R\approx 0.5$ $\mu $m for ice
plates at -15 C, so the width of the last molecular terrace is roughly $%
w\approx (aR)^{1/2}\approx 40a,$ where $a\approx 0.3$ nm is the size of a
water molecule. We might expect surface melting to be enhanced on such a
narrow terrace, owing to somewhat decreased molecular binding. Since surface
melting likely affects the step energy and thus $\sigma _{0}(T)$, it follows
that $\sigma _{0}$ may be lower on the thin crystal edge.

Looking at this from a different perspective, our measurements in Figure \ref%
{intrinsic} indicate that $\sigma _{0}$ generally decreases with increasing
temperature on both facets. This overall trend may result from a
\textquotedblleft softening\textquotedblright\ of the step edge at higher
temperatures, resulting from surface reconstruction relative to the simplest
one-molecule-wide step. The fact that $\beta \ll a\gamma ,$ where $\gamma $
is the surface energy, qualitatively supports this view \cite{intrinsic}. If
the step edge is further blurred on a thin terrace, then this would result
in the proposed SDAK effect.

The SDAK hypothesis is clearly quite speculative, without a firm theoretical
basis in molecular dynamics. This is necessarily the case, however, as there
is no real molecular theory of surface melting and ice surface dynamics, so
we cannot at present make quantitative statements about how $\sigma _{0}$
should depend on surface structure. The above basic physical considerations
suggest that the SDAK hypothesis is not unreasonable, and it is certainly
not disallowed by molecular dynamics considerations. Note also that the
growth rate of a thin crystal edge is largely determined by the attachment
kinetics on the final molecular terrace. The SDAK hypothesis requires only
that the molecular dynamics atop that last thin terrace be altered.

We also note that the SDAK hypothesis presents no conflict with the
Gibbs-Thomson phenomenon. The latter is a well-known effect by which the
equilibrium vapor pressure increases with increasing surface curvature,
following from simple surface energy considerations \cite{saito}. How the
step energy, and thus $\sigma _{0},$ becomes altered on a thin molecular
terrace is a more complex and somewhat orthogonal question. At the molecular
level, these phenomena may all be interconnected, but these connections are
not yet known from our present understanding of the ice surface structure
and dynamics.

For the present discussion we treat the SDAK model as a hypothesis to be
tested by observations. Our goal is then to examine whether this hypothesis
fits experimental data, and to use additional measurements and modeling to
better understand how structure-dependent attachment kinetics affects ice
crystal growth dynamics.

\begin{figure}[tb] 
  \centering
  \includegraphics[bb=0 0 3194 1431,width=6.07in,keepaspectratio]{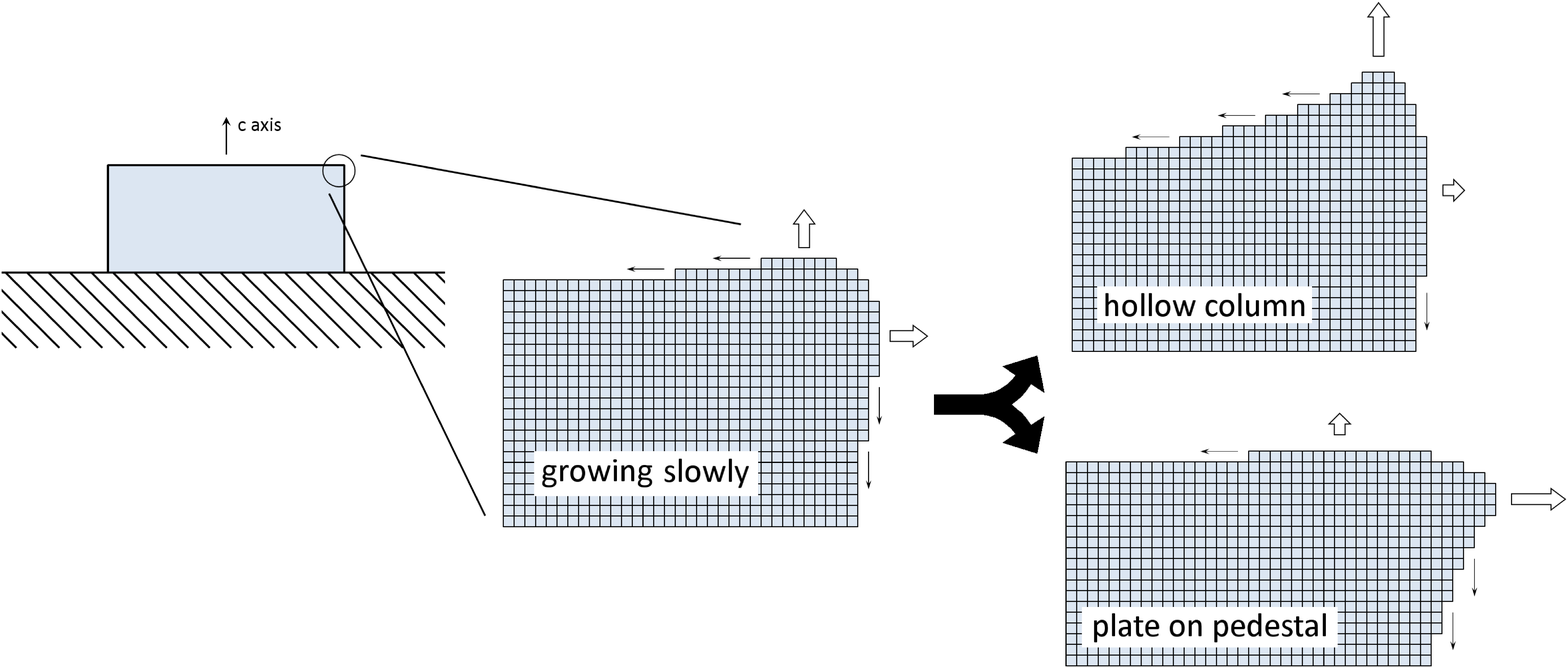}
  \caption{A schematic depiction of the
SDAK instability described in the text. The growth of the corner of a
faceted ice crystal prism (left) is dominated by the nucleation of terraces
on the basal and prism facets (center). If the nucleation rate increases as
the width of the top basal terrace decreases (top right), in keeping with
the SDAK model, then the accelerated growth narrows the basal surface,
accelerating the growth still more. The resulting positive feedback
generates the growth of a hollow columnar crystal. If the same SDAK effect
is more prevalent on the prism facet (lower right), then the instability
leads to the growth of a thin plate from the top edge of the prism.}
  \label{sdak}
\end{figure}

\subsection{An Edge-Enhancing SDAK\ Instability}

An important feature of the SDAK hypothesis is that it leads to an
edge-enhancing growth instability. The essential mechanism is that as a thin
edge begins to form, $\sigma _{0}$ decreases and thus further increases the
edge growth rate. The enhanced growth causes the edge to sharpen, which
again increases the growth rate. This positive feedback yields a growth
instability that enhances the formation of sharp edges.

To examine the instability in more detail, consider the growth of an
initially isometric prism on a substrate, depicted in Figure \ref{sdak}. We
assume the presence of an inert background gas surrounding the crystal, so
the growth is partially diffusion limited. If the crystal is growing slowly
(center diagram in the figure), then molecular terraces nucleate slowly near
the corner of the crystal, where the supersaturation is highest, and steps
propagate away from the corner. The corner itself is rounded from the
Gibbs-Thomson effect. For slow growth, this is essentially the standard
model of diffusion-limited faceted crystal growth, resulting in slightly
concave faceted surfaces.

Consider now the top terrace on either side of the growing corner. As the
supersaturation is increased, terraces nucleate more frequently, so the
average width of the top terrace decreases. As the SDAK effect reduces $%
\sigma _{0}$ on the narrower terraces, the nucleation rate increases and in
turn the more rapid growth further decreases the width of the top terrace.

At this point a competition occurs between growth on the basal and prism
facets, as shown in the pair of diagrams on the right side of Figure \ref%
{sdak}. If the SDAK effect preferentially reduces $\sigma _{0}$ on the basal
facet (top right diagram), then the basal growth is especially enhanced.
Because the growth is also diffusion-limited, the fast growth on the basal
facet depletes the water vapor supply from the nearby prism facet. This
decreases the nucleation rate on the prism facet, which causes the average
width of the last terrace to increase, which in turn increases $\sigma
_{0,prism}.$ The combined effect is that $\sigma _{0,prism}$ increases to
its intrinsic value while $\sigma _{0,basal}$ grows ever smaller as the
basal edge grows sharper. The final result is a hollow column morphology
with thin basal edges. Alternatively, the same instability could favor the
prism facets, as seen in the lower right diagram in Figure \ref{sdak}. In
this case a thin plate-like crystal would form on the isometric prism,
producing a \textquotedblleft plate-on-pedestal\textquotedblright\ (POP)
morphology, described in more detail below.

Note that the SDAK instability nicely explains the abrupt transitions
between plate-like and columnar growth seen in the morphology diagram.
Relatively small changes in the surface attachment kinetics with temperature
can be amplified via the SDAK instability to yield very substantial changes
in the final crystal morphologies.

Note also that the SDAK instability is essentially an extension of the
well-known Mullins-Sekerka instability in diffusion-limited growth \cite%
{mullins}. The latter is well known for producing dendritic branching during
solidification, but alone it does not automatically explain the formation of
thin plate-like or hollow columnar crystals. From our diffusion modeling of
crystal growth, we have found that these thin-edge morphologies require
strong anisotropies in the attachment kinetics -- namely $\alpha _{prism}\gg
\alpha _{basal}$ for thin plates or $\alpha _{prism}\ll \alpha _{basal}$ for
hollow columns. The SDAK instability provides a natural mechanism to
generate these strong anisotropies.

\section{Experimental Evidence for the SDAK Model}

The first piece of evidence supporting the SDAK hypothesis is simply the
disagreement between measurements of the intrinsic $\alpha (\sigma _{surf},T)
$ and the morphology diagram, as described above. The SDAK mechanism nicely
explains the formation of thin plates at -15 C, for example, which is
otherwise difficult to reconcile with the measured $\alpha _{basal}\left(
\sigma _{surf}\right) $ and $\alpha _{prism}\left( \sigma _{surf}\right) $
at that temperature. Another piece of supporting evidence is the observation
of abrupt morphological transitions between plate-like and columnar forms in
the morphology diagram, which again is naturally explained by the SDAK model.

These pieces of evidence provide only indirect support for the SDAK model,
however, based mainly on morphological observations. Quantitative modeling
of growth data provides a much stronger confirmation, as we demonstrate
below. Detailed measurements and diffusion modeling of growth behaviors at
-15 C fit the SDAK model well, while we see no easy way to explain the data
without invoking the SDAK effect.

\begin{figure}[htbp] 
  \centering
  \includegraphics[height=7.0in,keepaspectratio]{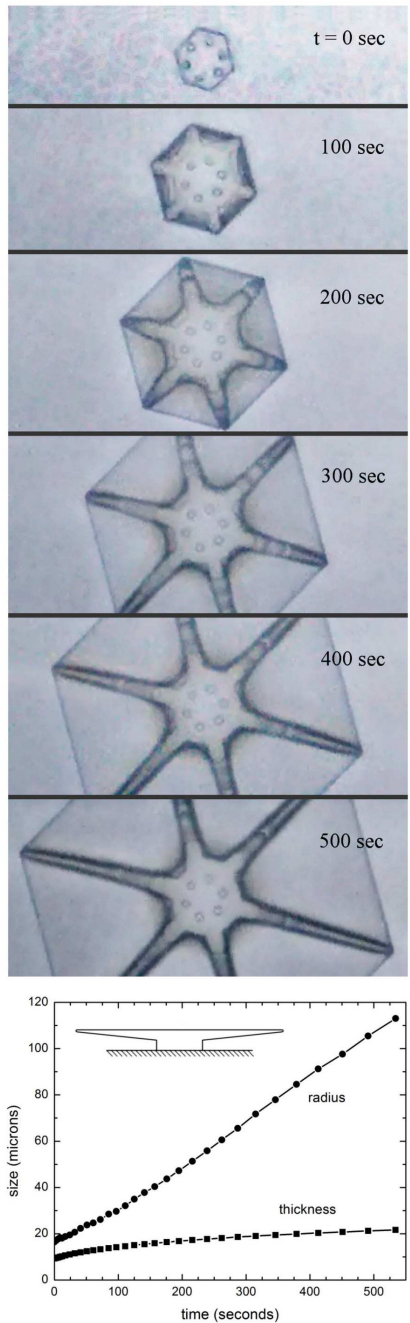}
  \caption{An example of \textquotedblleft
Plate-on-Pedestal\textquotedblright\ (POP) crystal growth at -15 C. The top
series of images shows the development of a thin sectored plate on top of an
initial hexagonal prism. The graph shows measurements of the plate radius
and central thickness as a function of time. The inset diagram in the graph
shows the inferred cross section of the crystal at $t=500$ seconds.}
  \label{popexample}
\end{figure}

\subsection{Plate-on-Pedestal Growth}

The clearest experimental evidence we have found to date supporting the SDAK
hypothesis comes from observations of what we call \textquotedblleft
plate-on-pedestal\textquotedblright\ (POP) crystal growth, and an example is
shown in Figure \ref{popexample}. The top image at $t=0$ in this figure
shows a small ice crystal with one basal facet resting on a sapphire
substrate. The initial morphology was essentially that of a simple hexagonal
prism, but with six small pits on one basal facet. These surface features
were trapped against the substrate, as evidenced by the fact that they
remained essentially unchanged during the subsequent growth of the crystal.

The substrate and test crystal in Figure \ref{popexample} lie at the bottom
of a small growth chamber. The top of the chamber is an ice reservoir (IR)
consisting of another sapphire surface covered with small ice crystals. The
IR provides a source of water vapor for the test crystal, and the
temperature difference $T_{IR}-T_{subst}$ determines the supersaturation far
from the test crystal, which we denote as $\sigma _{\infty }$. Additional
experimental details can be found in \cite{intrinsic}. In these experiments
the test crystal is small enough, and the ice thermal conductivity is high
enough, that the temperature of the test crystal is essentially equal to $%
T_{subst}.$

At $t=0$ in Figure \ref{popexample}, the supersaturation was increased from $%
\sigma _{\infty }=0$ to $\sigma _{\infty }=8$ percent, in an atmosphere of
air at a pressure of one bar. By $t=100$ seconds, a thin plate had begun
growing from the top edge of the prism, and this plate grew larger with
time. The ridges dividing the plate into six sectors formed on the underside
of the plate (i.e., the side of the plate nearest the substrate). Ridging of
this kind is commonly seen in the growth of plate-like ice crystals from
water vapor \cite{libbrechtreview}. The graph in Figure \ref{popexample}
shows the radius (here defined as half the distance between opposite prism
facets) and thickness (equal to the distance from the substrate to the
center of the upper basal facet) of the crystal as a function of time. The
thickness was measured using optical interferometry as described in \cite%
{intrinsic}, while the radius was determined from the optical images. The
inset in the graph shows the approximate cross section of the crystal at $%
t=500$ seconds. The detailed morphology of the ridging was not measured, and
the top surface of the plate may be slightly conical in shape, and thus not
precisely flat as indicated in the inset cross-section diagram.

The POP morphology shown in Figure \ref{popexample} was observed at -15 C as
long as $\sigma _{\infty }$ was above about 4-5 percent (depending on the
initial crystal radius and thickness). For lower $\sigma _{\infty }$, simple
prisms grew and the POP\ morphology did not develop. For these low-$\sigma
_{\infty }$ crystals, the prism growth was additionally influenced by
substrate interactions. The presence of chemical residue on the substrate
(even after thorough cleaning) often lowered the nucleation barrier on
facets intersecting the substrate, resulting in growth rates that were
somewhat higher than one would expect from the intrinsic $\alpha $
measurements in Figure \ref{intrinsic}. This effect is discussed further in 
\cite{intrinsic}, but it played only a minor role in the current
measurements.

Note that the POP\ morphology also does not appear when the background
pressure is low. The measurements yielding the results in Figure \ref%
{intrinsic} were taken at pressures below 30 mbar, where diffusion is rapid
and the growth is mainly kinetics limited. In general the crystal morphology
is more complex at higher pressures and for larger crystals \cite{gonda}, a
fact that follows from scaling relationships in solutions to the diffusion
equation \cite{cakgl}.

\begin{figure}[htb] 
  \centering
  \includegraphics[height=3.5in,keepaspectratio]{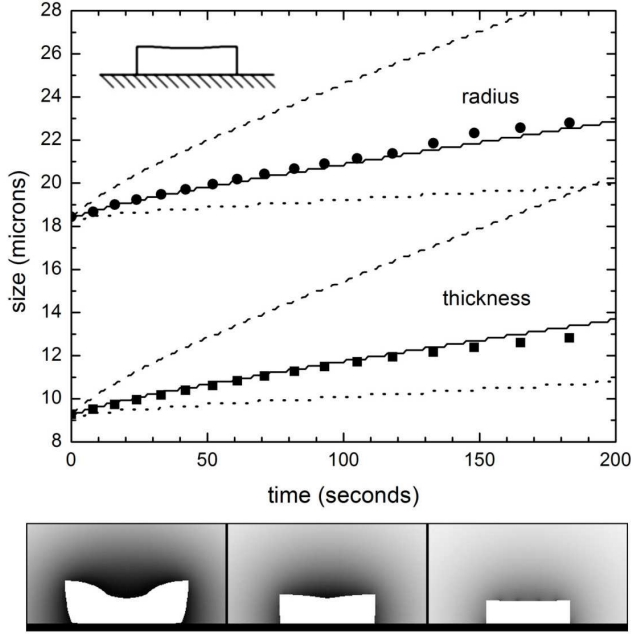}
  \caption{Modeling the growth of a
relatively simple prism crystal. Data points show measurements of the radius
and central thickness of the crystal as a function of time, and the applied
supersaturation was $\protect\sigma _{\infty }=3$ percent. Lines show
numerical models of the growth using $\protect\alpha =\exp (-\protect\sigma %
_{0}/\protect\sigma _{surf})$ with $\protect\sigma _{0,basal}=\protect\sigma %
_{0,prism}=2$ percent. The dashed lines, solid lines, and dotted lines
indicate models with $\protect\sigma _{\infty }=3$, 1.5, and 0.75 percent,
respectively. The inset in the graph shows the inferred cross section of the
crystal at 200 seconds. The images below the graph show cross sections
calculated in the three models at 200 seconds, with $\protect\sigma _{\infty
}$ decreasing from left to right. In these images the crystal is shown in
white, and the brightness of the surrounding area is proportional to the
supersaturation field.}
  \label{3percent1}
\end{figure}

Before considering a quantitative analysis of the growth of POP crystals, we
first examine the growth of simple prism crystals, and an example is shown
in Figure \ref{3percent1}. In this figure, the data points show measurements
of the radius and thickness of the crystal as a function of time, similar to
the data shown in Figure \ref{popexample}, also taken in air at a pressure
of one bar. For this crystal, however, the supersaturation was only $\sigma
_{\infty }=3$ percent, so the growth was slower and the POP morphology did
not develop. The crystal showed a slight basal hollowing at $t=200$ seconds,
as shown in the cross-section diagram inset in the graph.

Diffusion modeling was performed using the cylindrically symmetrical
cellular automaton model described in \cite{cakgl}. The outer radial surface
in this model corresponds to a single prism \textquotedblleft
facet,\textquotedblright\ which we found to be an adequate geometrical
approximation to the six prism facets of a simple hexagonal prism. In
addition to simple prisms, hollow columns and POP morphologies can also be
effectively modeled with this cylindrically symmetrical method.

For the models shown in Figure \ref{3percent1} and in subsequent figures
below, the outer boundary with $\sigma =\sigma _{\infty }$ was set at $%
r_{outer}=105$ $\mu m$ and $z_{outer}=75$ $\mu m.$ The diffusion equation
was solved in the space surrounding the crystal to determine the
supersaturation field $\sigma (r,z),$ with input $\alpha _{basal}\left(
\sigma _{surf}\right) $ and $\alpha _{prism}\left( \sigma _{surf}\right) $
for the facet surfaces, assuming $\alpha =1$ at kink sites on the surface 
\cite{cakgl}. The quantitative accuracy of the modeling software was
thoroughly tested using analytic solutions for the growth of spherical and
infinite cylindrical crystals.

We believe we have a reasonable understanding of the factor-of-two
discrepancy between the data taken at $\sigma _{\infty }=3$ percent and the
best-fit model with $\sigma _{\infty }=1.5$ percent (see Figure \ref%
{3percent1}). About half of the discrepancy arises from the model itself.
Part comes from the fact that the outer boundaries in the model are fairly
close to the crystal, which causes the model to grow more rapidly than it
would if the boundaries were as far away as in the experiment. Another part
comes from our choice for the adaptive time steps in the model \cite{cakgl}.
With longer time steps, the supersaturation field does not have time to
relax fully as the crystal grows, again causing the model crystal to grow
too fast. Both these effects were investigated using analytic solutions, and
a compromise between modeling accuracy and speed was made.

Even using shorter time steps and more distant outer boundaries, however, we
could not obtain a good fit to the data using a model input of $\sigma
_{\infty }=3$ percent. We believe the remainder of the discrepancy was
caused by distant neighboring crystals in the experiment. These additional
crystals act as water vapor sinks, reducing the supersaturation around the
crystal by a small amount, and this effect apparently caused an effective $%
\sigma _{\infty }$ that was roughly 30 percent lower than that calculated
from $T_{IR}-T_{subst}$. We believe that the combined effects of these
modeling and experimental systematics reasonably explain the factor of two
discrepancy between the applied $\sigma _{\infty }$ in the experiment and
the best fit model.

With the crystal in Figure \ref{3percent1} and others, we found that a good
modeling strategy was to adjust the assumed $\sigma _{\infty }$ values in
the model to match the overall growth rates observed. This seemed to
adequately fit for both the modeling and experimental systematics in $\sigma
_{\infty }$ without adverse effects. Typically the best fit $\sigma _{\infty
}$ value was about a factor of two below the experimental value determined
from $T_{IR}-T_{subst}$, with some crystal-to-crystal variation reflecting
differences in experimental conditions.

\begin{figure}[htb] 
  \centering
  \includegraphics[height=3.5in,keepaspectratio]{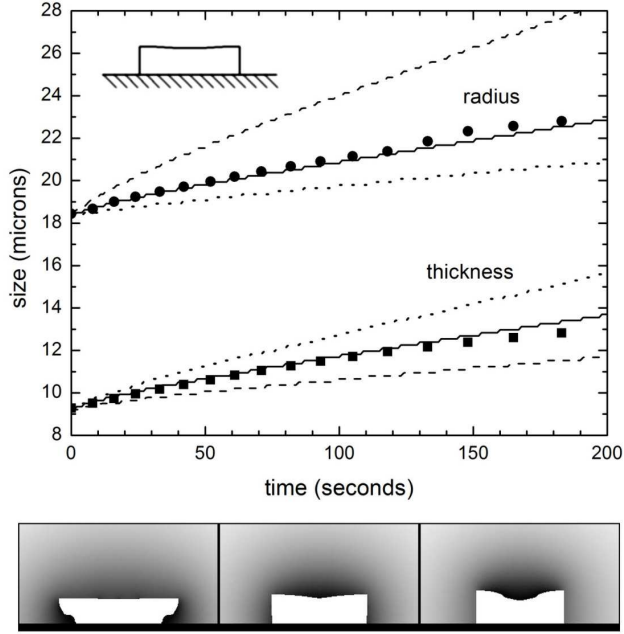}
  \caption{Additional modeling of the same
low-$\protect\sigma _{\infty }$ crystal shown in Figure \protect\ref%
{3percent1}. In these models we fixed $\protect\sigma _{\infty }=1.5$
percent while keeping $\protect\sigma _{0,basal}=2$ percent. The dashed,
solid, and dotted lines show models with $\protect\sigma _{0,prism}=1,$ 2,
and 3 percent, respectively. Images below the graph again show model cross
sections, with $\protect\sigma _{0,prism}$ increasing from left to right.}
  \label{3percent2}
\end{figure}

Figure \ref{3percent2} shows additional modeling of the same low-$\sigma
_{\infty }$ crystal shown in Figure \ref{3percent1}. Here we see that a
model with $\sigma _{0,prism}=3$ percent -- the preferred value from the
intrinsic data in Figure \ref{intrinsic} -- yields prism growth that is too
slow and basal growth that is too fast, along with too much hollowing in the
basal facet. In contrast, the model with $\sigma _{0,prism}=1$ percent
yields prism growth that is too fast and basal growth that is too slow,
along with the beginnings of a POP-like morphology. The model with $\sigma
_{0,prism}=2$ percent matches the data reasonably well. The fact that the
best fit value of $\sigma _{0,prism}=2$ percent does not agree with
intrinsic growth measurements was not unexpected, given the known influence
of substrate interactions mentioned above.

Our conclusions from this modeling exercise include: 1) our numerical
methods produce adequate quantitative models of the growth of simple prism
crystals, 2) adjusting $\sigma _{\infty }$ is a reasonable strategy to fit
the overall growth rates, 3) a value of $\sigma _{0,basal}\approx 2$
percent, consistent with the intrinsic growth rate shown in Figure \ref%
{intrinsic}, gives a reasonable fit to the data, and 4) a value of $\sigma
_{0,prism}\approx 2$ percent, likely decreased from its intrinsic value by
substrate interactions, is needed to reproduce the growth of this low-$%
\sigma _{\infty }$ crystal. The significance of this latter result becomes
apparent when we compare it with models of the growth of a POP crystal.

\begin{figure}[htb] 
  \centering
  \includegraphics[height=3.5in,keepaspectratio]{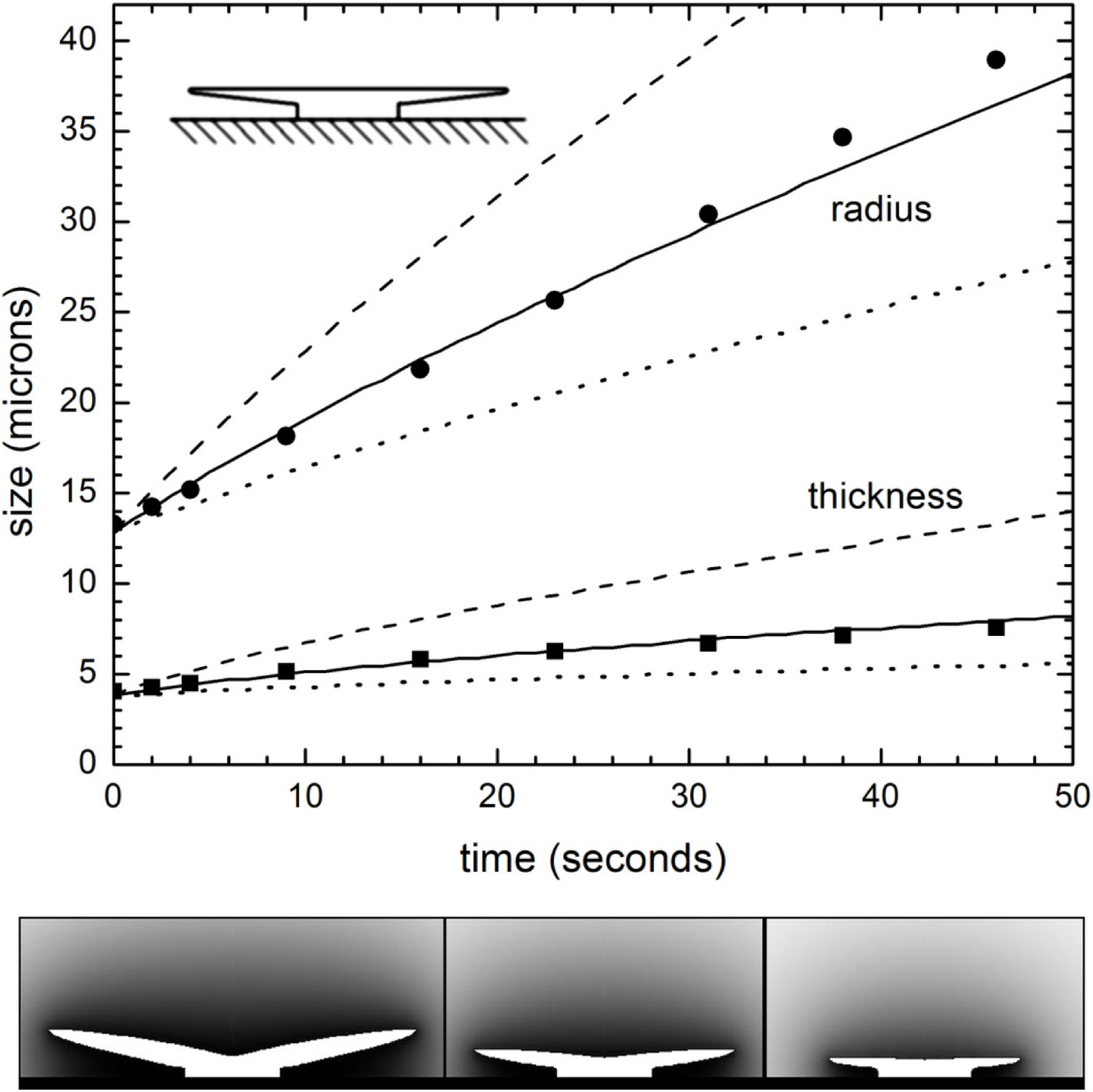}
  \caption{An example of the growth of a POP\
crystal with an applied supersaturation of $\protect\sigma _{\infty }=10$
percent. This crystal began at $t=0$ with a simple prism morphology, then
quickly developed a POP morphology. The points in the graph show the
measured plate radius and central crystal thickness as a function of time.
The inset diagram shows the inferred crystal cross section at 50 seconds.
The dashed, solid, and dotted lines show models with $\protect\sigma %
_{\infty }=10,$ 5, and 2.5 percent, respectively, using $\protect\sigma %
_{0,basal}=2$ percent and $\protect\sigma _{0,prism}=0.3$ percent. The
images below the graph show the corresponding model cross sections at 50
seconds.}
  \label{10percent1}
\end{figure}

Figure \ref{10percent1} shows the growth of a second example crystal,
similar to the results in Figure \ref{3percent1} but with a higher applied
supersaturation of $\sigma _{\infty }=10$ percent. Modeling proceeded as
before, and we see that again the overall growth scales with $\sigma
_{\infty },$ giving a best fit $\sigma _{\infty }\ $that is once more about
a factor of two smaller than the experimentally calculated supersaturation.
And again we see that adjusting $\sigma _{\infty }$ in the models allows us
to reasonably fit the overall morphology as well as the quantitative growth
measurements.

\begin{figure}[htb] 
  \centering
  \includegraphics[height=3.5in,keepaspectratio]{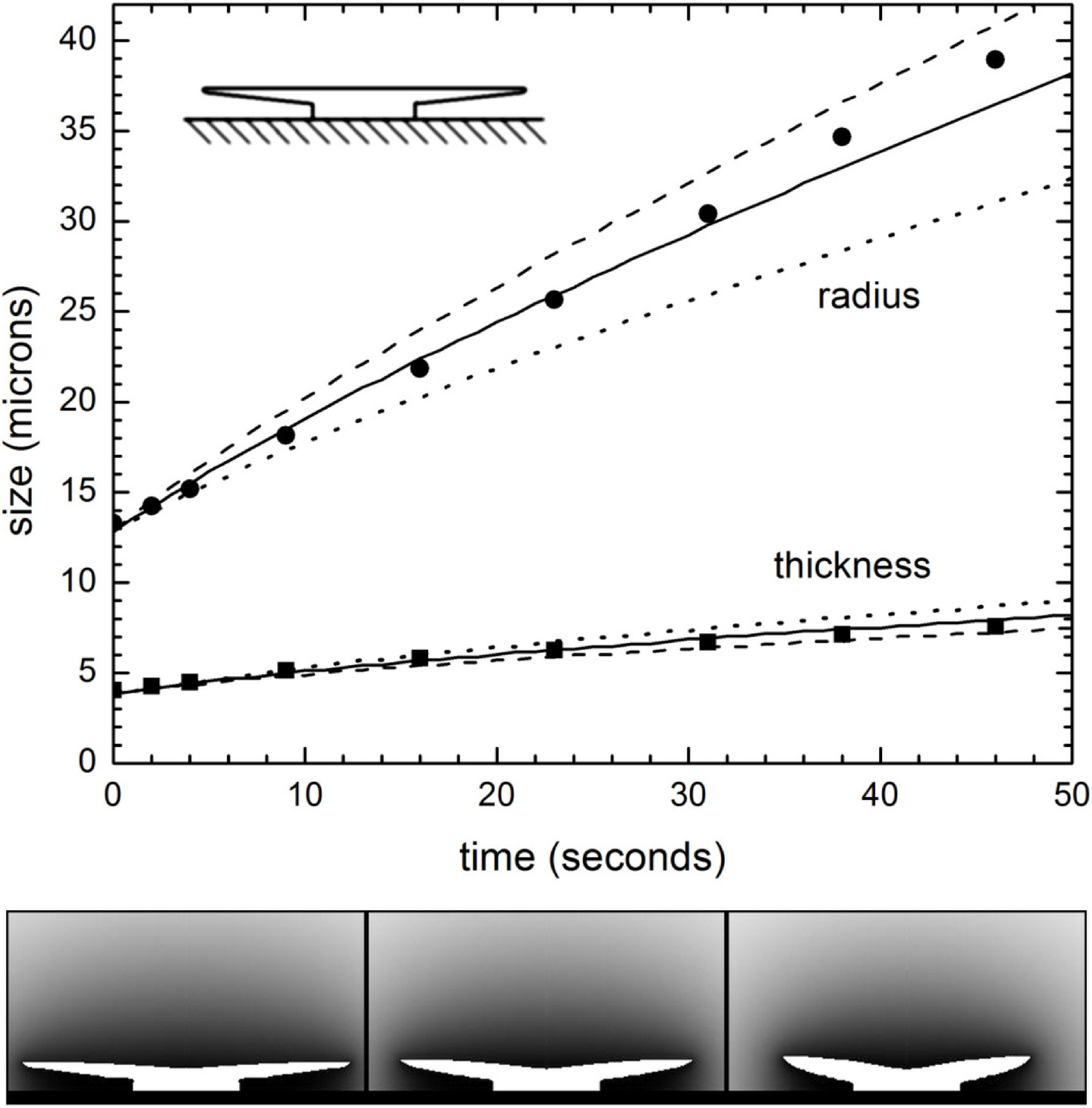}
  \caption{Additional models of the same
high-$\protect\sigma _{\infty }$ shown in Figure \protect\ref{10percent2}.
For these models we fixed $\protect\sigma _{\infty }=5$ percent and $\protect%
\sigma _{0,basal}=2$ percent. The dashed, solid, and dotted lines show
models with $\protect\sigma _{0,prism}=0.15,$ 0.3, and 0.6 percent,
respectively.}
  \label{10percent2}
\end{figure}

Figure \ref{10percent2} shows another comparison of three models, this time
keeping $\sigma _{\infty }=5$ percent and $\sigma _{0,basal}=2$ percent,
while varying $\sigma _{0,prism}$. Here we see that the basal growth depends
only weakly on our choice for $\sigma _{0,prism}$ (over this range), while
changing $\sigma _{0,prism}$ affects the radial growth rate and the overall
POP morphology.

Note that we did not attempt to construct a numerical model that included
the full SDAK effect. Our models were more basic in that they assumed
attachment coefficient functions $\alpha _{basal}\left( \sigma
_{surf}\right) $ and $\alpha _{prism}\left( \sigma _{surf}\right) $ that did
not depend on the structure of the crystal. This was appropriate for low-$%
\sigma _{\infty }$ crystals, where the SDAK effect was absent, but the
models also worked quite well for describing POP\ crystals. The thin edge
developed very quickly on the high-$\sigma _{\infty }$ crystal in Figures %
\ref{10percent1} and \ref{10percent2}, and the subsequent plate growth was
determined mainly by $\alpha _{prism}$ on the edge. The only other prism
facets present were on the pedestal, and these surfaces were so heavily
shielded that their growth was slow regardless of $\alpha _{prism}.$ For the
specific POP geometries we observed, therefore, our basic models yielded
essentially the same quantitative growth behavior as we would have obtained
with a full SDAK model.

\subsection{Conclusions}

We draw several conclusions from a comparison of the growth of these two
crystals: 1) changing $\sigma _{\infty }$ in the models changes the overall
growth rates of the crystals, as one would expect. Adjusting the model $%
\sigma _{\infty }$ to be about half the experimental $\sigma _{\infty }$
gives a good fit to the data, and the need for this factor of two is
reasonably well understood; 2) a value of $\sigma _{0,basal}=2$ percent
gives a reasonable fit to all the data, and this value is consistent with
the intrinsic growth measurements in Figure \ref{intrinsic}; and 3) the low-$%
\sigma _{\infty }$ prism crystal requires $\sigma _{0,prism}\approx 2$
percent to fit the data, while the high-$\sigma _{\infty }$ POP crystal
requires $\sigma _{0,prism}\approx 0.3$ percent.

We observed numerous other crystals in addition to these two examples, and
the overall growth behaviors are consistent, leading to essentially the same
conclusions. Each observed crystal had a different initial radius and
thickness, plus the supersaturation and substrate interactions differed
slightly from run to run. For this reason we found it advantageous to
present case studies of these two example crystals.

The most significant conclusion from these data is that it was clearly not
possible to model both crystals adequately using a single-valued function $%
\alpha _{prism}(\sigma _{surf})$ for the prism attachment coefficient. This
is perhaps most clearly seen by examining the prism growth velocities as a
function of $\sigma _{surf},$ as the latter can be obtained from the
best-fit models. For the low-$\sigma _{\infty }$ crystal at $t=200$ seconds,
we found a maximum value of $\sigma _{surf}\approx 0.51$ percent on the
prism facet, yielding a growth rate of $v_{prism}\approx 20$ nm/sec. For the
high-$\sigma _{\infty }$ crystal at $t=50$ seconds, we found a maximum $%
\sigma _{surf}\approx 0.67$ percent on the edge of the plate, yielding a
growth rate of $v_{prism}\approx 560$ nm/sec. Although the supersaturations
at the prism surfaces in these crystals differed only slightly, the thin
edge of the POP\ crystal grew nearly 30 times faster.

The impact of this result becomes apparent when one tries to produce
quantitative models of the growth behavior without invoking the SDAK
hypothesis. We examined a number of different crystals and explored
different modeling strategies, but in the end the conclusion was robust and
clear -- the attachment kinetics on the edge of the thin plate was much
faster than on the edge of a thick plate. These measurements strongly
supports the SDAK hypothesis, and we could not find another reasonable
hypothesis that would quantitatively explain the data.

\subsection{SDAK or VDAK?}

One alternative hypothesis we explored in some detail was \textit{%
velocity-dependent attachment kinetics} (VDAK), which in this case assumes $%
\alpha _{prism}$ can depend on the surface growth velocity. The VDAK
hypothesis is motivated by kinetic roughening \cite{saito}, and we consider
VDAK as a generalization of this phenomenon. In a nutshell, kinetic
roughening describes a process where sufficiently fast crystal growth causes
the growing surface (normally faceted) to become disordered at the molecular
scale, with a resulting jump in the attachment kinetics. For the case of POP
crystal growth, assuming a VDAK hypothesis implies that the edge of the thin
plate is growing sufficiently fast to produce a partial roughening, and thus
an increase in $\alpha _{prism}$ on the edge.

The VDAK hypothesis could perhaps explain the POP data above, but it also
leads to other testable predictions that we found were not verified by
experiment. The first prediction would be a jump in the growth rate once a
critical growth velocity is exceeded. This jump would indicate a VDAK
transition, and the jump should occur on large faceted surfaces as well as
on thin edges. And, importantly, the jump should occur both with or without
a background gas.

A second prediction is that the VDAK hypothesis should lead to a different
kind of growth instability. When $\sigma _{surf}$ is increased and the
attachment kinetics starts to increase from the VDAK effect, this should in
turn increase the growth velocity for the same $\sigma _{surf}$, thus
further increasing the attachment kinetics. This positive feedback would
likely result in a rapid jump to fully roughened growth with $\alpha \approx
1.$ Again we would expect this instability to occur on large facets or thin
edges, with or without a background gas.

Thus the VDAK instability should result in a hysteresis behavior in growth
measurements. Increasing and then decreasing $\sigma _{surf}$ should lead to
a growth velocity $v\left( \sigma _{surf}\right) $ that is a double-valued
function of $\sigma _{surf}$ -- low for increasing $\sigma _{surf}$ and
higher for decreasing $\sigma _{surf},$ as long as the critical velocity is
exceeded at some point.

\begin{figure}[htbp] 
  \centering
  \includegraphics[height=5.8in,keepaspectratio]{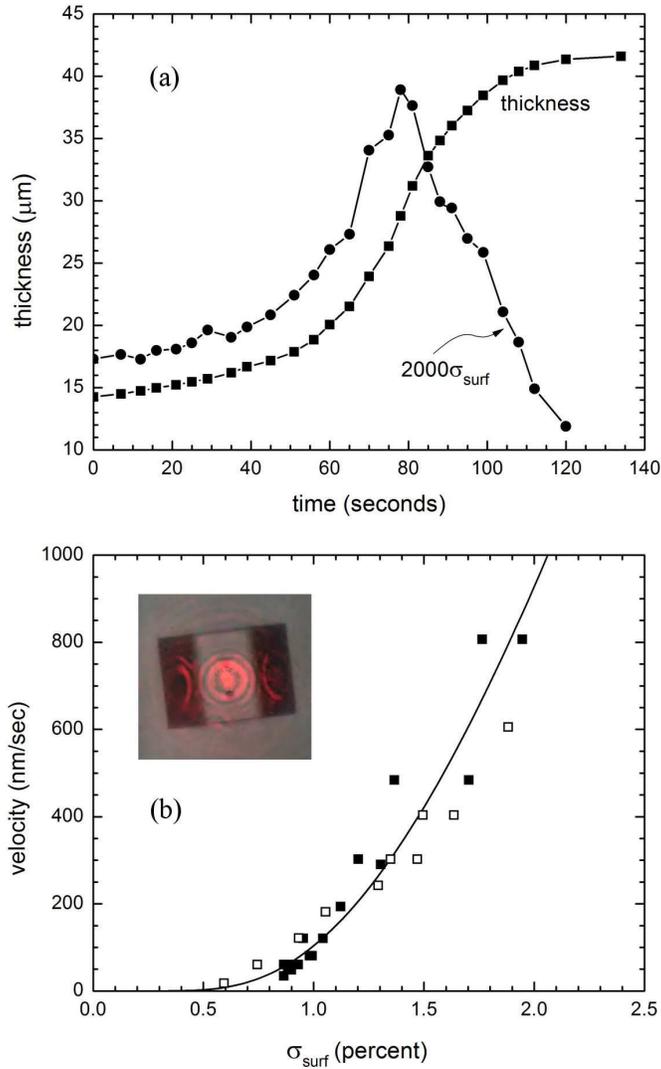}
  \caption{A measurement of prism facet
growth at -15 C in a background of air at a pressure of 20 mbar. In (a) we
present measurements of the crystal thickness (defined here as the distance
from the substrate to the top prism facet) as a function of time. The
supersaturation at the surface is also plotted, scaled to 2000$\protect%
\sigma _{surf}$ to fit on the graph. In (b) the measured prism growth
velocity is plotted as a function of $\protect\sigma _{surf}.$ Solid points
give $v(\protect\sigma _{surf})$ as $\protect\sigma _{surf}$ was being
increased; open points show $v(\protect\sigma _{surf})$ as $\protect\sigma %
_{surf}$ was being decreased. The line shows $v=\protect\alpha v_{kin}%
\protect\sigma _{surf}$ with $\protect\alpha =\exp (-\protect\sigma _{0}/%
\protect\sigma _{surf})$ and $\protect\sigma _{0}=3$ percent. The inset
image in (b) shows the test crystal at the end of the run. Oscillations in
the brightness of the laser spot were used to interferometrically measure
the crystal thickness. In both these plots, $\protect\sigma _{surf}$
includes a correction to remove residual diffusion effects. A description of
this correction, along with additional experimental details, can be found in 
\protect\cite{intrinsic}.}
  \label{vdak}
\end{figure}

We tested these predictions by measuring prism facet growth as a function of 
$\sigma _{surf}$ at $T=-15$ C with a background pressure of air at 20 mbar,
in a fashion similar to the experiments described in \cite{intrinsic}. The
main changes were that we pushed the growth to much higher peak velocities,
and we quickly ramped $\sigma _{surf}$ up and then down, measuring the prism
growth rate in the process. An example crystal from this set of measurements
is shown in Figure \ref{vdak}.

Graph (b) in Figure \ref{vdak} shows several significant things:\ 1) we
observed no jump in growth as a function of $\sigma _{surf},$ even though
the peak velocity was over 800 nm/sec, faster than the growth velocities
observed in the POP crystals above; 2) the observed $v(\sigma _{surf})$ was
consistent with the intrinsic growth measurements shown in Figure \ref%
{intrinsic}; and 3) we observed no hysteresis in $v(\sigma _{surf})$ -- the
measured velocities with $\sigma _{surf}$ increasing were not significantly
different from the velocities with $\sigma _{surf}$ decreasing.

The data shown in Figure \ref{vdak} show no evidence of any of the expected
VDAK effects, so we conclude that the VDAK hypothesis is refuted by these
data. If a VDAK effect does exist, it seems to play no significant role in
the prism facet growth at -15 C. Note that again we observed more than the
single crystal shown in Figure \ref{vdak}, and the overall conclusions were
quite robust.

\begin{figure}[tb] 
  \centering
  \includegraphics[height=3.8in,keepaspectratio]{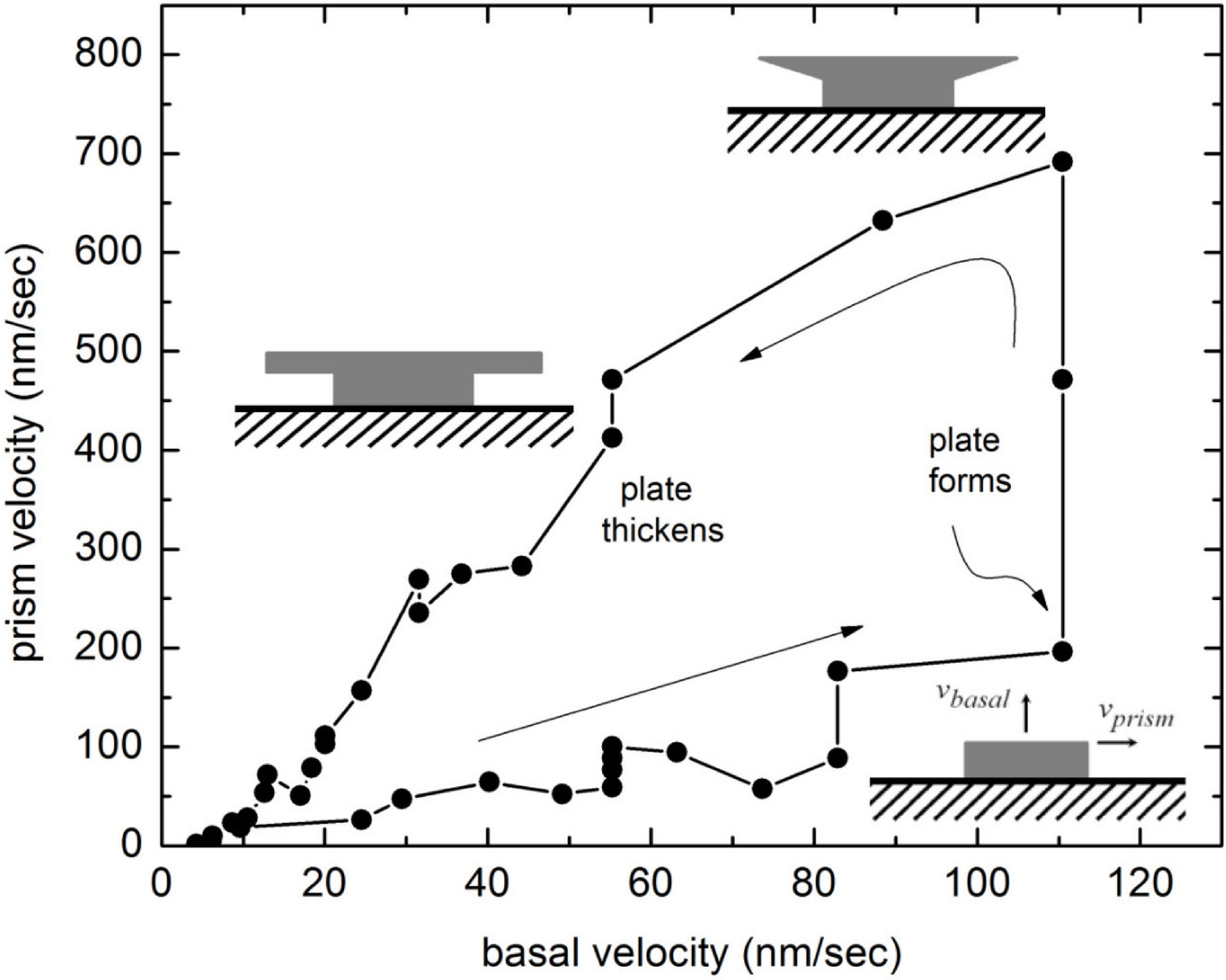}
  \caption{An example showing hysteresis in
POP crystal growth, here plotting $v_{prism}$ versus $v_{basal}$ as the
crystal grew. The inital growth (lower track in the measured points)
exhibited a simple prism morphology (lower right inset diagram) as $\protect%
\sigma _{\infty }$ was slowly increased, with $v_{prism}\approx
1.5v_{basal}. $ When the plate formed, $v_{prism}$ abruptly increased while $%
v_{basal}$ remained constant (top inset diagram). Beginning at the $%
v_{prism} $ peak, $\protect\sigma _{\infty }$ was slowly decreased, causing
both $v_{prism}$ and $v_{basal}$ to decrease as the plate edge thickened
(left inset diagram).}
  \label{pophysteresis}
\end{figure}

\subsection{Hysteresis in Plate-on-Pedestal Growth}

Although we did not observe any hysteresis behavior that could be ascribed
to VDAK effects, we did observe a clear hysteresis behavior in the growth of
POP crystals, and an example from this set of measurements is shown in
Figure \ref{pophysteresis}. These data were acquired by slowly increasing $%
\sigma _{\infty }$ while observing the development of a POP\ crystal at -15
C in a background of air at a pressure of one bar. Additional experimental
details, and additional observations, can be found in \cite{sdak2}.

In Figure \ref{pophysteresis} we see that the initial growth of this
crystal, while $\sigma _{\infty }$ was small and slowly increasing, produced
a nearly isometric simple prism. In this case the prism growth velocity, $%
v_{prism},$ was roughly 1.5 times the basal growth velocity, $v_{basal}.$
This factor was likely affected by substrate interactions that influenced
the prism growth. As $v_{prism}$ approached 200 nm/sec, the POP morphology
began to form. This event in turn caused $v_{prism}$ to jump to about 700
nm/sec while $v_{basal}$ remained essentially constant.

Once the plate had fully formed, $\sigma _{\infty }$ was then slowly
decreased with time. This caused both $v_{basal}$ and $v_{prism}$ to
decrease, and it caused the edge of the plate to slowly thicken. However,
the growth in the $v_{basal}$-$v_{prism}$ plane did not track back down its
original path. Instead the prism growth remained high, with $%
v_{prism}\approx 8v_{basal}.$

Simple diffusion modeling with fixed $\alpha _{basal}\left( \sigma
_{surf}\right) $ and $\alpha _{prism}\left( \sigma _{surf}\right) $ cannot
reproduce this hysteresis behavior, but it is easily explained with the SDAK
model. During the initial stages of growth, when $\sigma _{\infty }$ was
fairly low, the crystal morphology was that of a simple prism, so there was
no SDAK effect. When the growth passed a threshold, the SDAK instability
caused a thin plate to form on the prism, and the accompanying decrease in $%
\sigma _{0,prism}$ resulted in a jump in $v_{prism}.$ The supersaturation
near the crystal surface remained approximately constant during this
process, as evidenced by the fact that $v_{basal}$ did not change
appreciably during the jump in $v_{prism}$.

As $\sigma _{\infty }$ was subsequently reduced, the SDAK instability began
to lose its strength, causing the plate to thicken. As the plate slowly
thickened, $\sigma _{0,prism}$ slowly increased back to its intrinsic value.
The initial jump in $v_{prism}$ was rapid because it took little time for
the SDAK instability to sharpen the edge of the plate. The subsequent
decrease in $v_{prism}$ took longer because it took more time for the plate
to thicken substantially.

The hysteresis behavior shown in Figure \ref{pophysteresis} is nicely
explained by the SDAK instability, so it provides additional supporting
evidence for the SDAK hypothesis. However the experiment was somewhat
complex and difficult to model accurately, while the underlying growth
behavior is essentially the same as with other POP crystals. Therefore we
found that the constant-$\sigma _{\infty }$ measurements described above
provided better quantitative support for the SDAK hypothesis.

\subsection{Additional SDAK Support}

The SDAK instability is most clearly seen in ice plate growth at -15 C, and
to date we have not mapped out the growth dynamics at other temperatures.
Nevertheless, quantitative modeling of the growth of ice crystals in free
fall \cite{freefall, fukuta1} does suggest that SDAK effects are also
present in the growth of columnar crystals near -5 C and in plate growth at
-2 C. Thus our preliminary investigations suggest that SDAK effects in ice
growth are substantial over a broad temperature range.

The SDAK instability may also explain the recent observations of
fast-growing needle-like structures alongside faceted crystals at -5 C
reported by Knight \cite{knight}. Here again, the simultaneous occurrence of
these quite different morphologies is a natural result of the SDAK
instability, but is otherwise difficult to explain.

\section{Discussion}

Progress toward understanding the formation of complex structures during
solidification has generally been hard won, as numerous physical processes
are involved over many length scales. Creating even qualitative models of
the subtle many-body dynamics governing crystal growth is difficult, and
realizing accurate numerical models that allow quantitative comparison with
experimental data remains a significant challenge.

It has been known since the mid-1960s, for example, that branched structures
arise from the Mullins-Sekerka instability during diffusion-limited growth.
Producing a quantitative model of this process required substantial
theoretical effort, however, culminating in the development of solvability
theory during the 1980s \cite{dendrites, brener}. With this we learned that
the overall branching scale is set by seemingly minor anisotropies in the
surface dynamics. The surface energy anisotropy plays the key role in the
case of solidification from the liquid phase, while for solidification from
gaseous precursors the anisotropy in the surface attachment kinetics is
generally the more dominant factor \cite{libbrechtreview}.

Numerical models of diffusion-limited growth were developed in conjunction
with solvability theory, including front-tracking and phase-field
techniques. These have enjoyed reasonable success in reproducing the salient
features observed in many liquid systems aimed at understanding
metallurgical solidification processes. These same numerical techniques have
been generally less successful modeling growth from the vapor phase,
however, including ice crystal growth, when the resulting structures are
both faceted and branched. For these systems, diffusion models derived from
cellular automata have proven more successful in reproducing observed
structures \cite{gg}. Understanding the differences in these numerical
modeling techniques, and producing quantitative models with more accurate
surface physics, remains an area of current research.

Analysis of diffusion-limited growth using these theoretical tools has
yielded numerous insights into the dynamics of structure formation. For the
case of ice, the increase in morphological complexity with increasing
supersaturation is reasonably well understood at a qualitative level,
although quantitative details are still lacking. Solvability theory nicely
explains how the tip velocity of a growing dendritic structure depends
linearly on supersaturation for solidification from vapor, while a quadratic
dependence on undercooling is typical for growth from the liquid phase \cite%
{libbrechtreview}. Furthermore, scaling relationships in diffusion growth
models provide an explanation for the increase in structural complexity that
accompanies decreasing diffusion rates \cite{cakgl}.

In spite of these numerous successes in growth modeling, many puzzles still
remain in understanding the detailed physics underlying the surface
attachment kinetics, especially in growth from the vapor phase when
attachment kinetics plays a substantial role. In the case of ice growth, the
problem is made considerably more difficult by the presence of surface
melting, since it is still largely unknown how this phenomenon influences
crystal growth dynamics.

Several experiments from recent years have indicated that faceted ice growth
is well described by a dislocation-free layer-nucleation model (\cite{nelson}%
, and references therein). Quantitative measurements have further shown that
the attachment coefficients on the principal facets can be parameterized
using $\alpha (\sigma )=A\exp (-\sigma _{0}/\sigma )$, with the measured $%
A(T)$ and $\sigma _{0}(T)$ shown in Figure 1. Classical nucleation theory
then allows one to convert $\sigma _{0}(T),$ a complex dynamical quantity,
into the step energy $\beta (T),$ a more basic equilibrium property of the
ice surface.

With the experimental and modeling results presented above, we have added a
new twist to our understanding of crystal growth dynamics, namely
structure-dependent attachment kinetics and the SDAK\ instability. Our data
and modeling indicate that the SDAK effect is a necessary ingredient for
reproducing the observed growth behavior of ice from water vapor at -15 C,
and it appears likely that the formation of columnar structures near -5 C
and plate-like forms near -2 C are also influenced by this phenomenon. We
believe, therefore, that the SDAK model described above represents a
substantial step forward in explaining the overall organization in the
Nakaya morphology diagram. It appears promising that additional features in
the model will emerge from future targeted experimental studies.

Ongoing studies of ice growth from water vapor are revealing ever more
features in an already rich phenomenology, and our list of observed growth
instabilities now includes the Mullins-Sekera instability, the SDAK
instability, and the electric growth instability \cite{electric}. To these
we add the rather complex behavior seen in the attachment kinetics, notably
different on the two principal facets, and it becomes clear that this
particular system still has much to teach us about the subtle interplay of
different physical processes governing crystal growth dynamics.

Experimental and theoretical efforts focusing on structure formation in ice
over the past several decades have continually pushed the remaining
frontiers ever closer to the molecular scale. The SDAK instability, the
parameterization of the attachment coefficients for the principal facets, as
well as the properties of surface melting, all arise from the detailed
molecular dynamics at the crystal surface. Since much progress has been made
recently in molecular dynamics simulations of the ice surface \cite%
{moldymice1, moldymice2}, it appears promising that additional
investigations along these lines may reveal new insights into ice growth
behavior, and especially why the ice surface properties vary with
temperature as they do. How these advances apply to other crystal systems,
and to our understanding of surface molecular dynamics in general, remains
to be seen.

\end{document}